\documentclass[12pt]{iopart}

\usepackage{iopams}
\usepackage{graphicx}
\usepackage{color}

\newcommand{\B}[1]{\mathbf{#1}}
\newcommand{\pol}{\sigma}

\newcommand{\text}{\mathrm}
\begin{document}

\title{Revealing short-range non-Newtonian gravity through Casimir-Polder shielding}

\author{Robert Bennett}

\address{Physikalisches Institut, Albert-Ludwigs-Universit\"{a}t Freiburg, Hermann-Herder-Str. 4 Freiburg 79104, Germany}
\address{Freiburg Institute for Advanced Studies, Albert-Ludwigs-Universit\"at Freiburg, Albertstra{\ss}e 19, 79104 Freiburg, Germany}
\ead{robert.bennett@physik.uni-freiburg.de}

\author{D. H. J. O'Dell}

\address{Department of Physics and Astronomy, McMaster University, 1280 Main St.\ W., Hamilton, ON, L8S 4M1, Canada}

\begin{abstract}
We carry out a realistic, yet simple, calculation of the Casimir-Polder interaction in the presence of a metallic shield in order to aid the design of experiments to test non-Newtonian gravity. In particular, we consider a rubidium atom near a movable silicon slab with a gold film in between. We show that by moving the slab to various distances and making precise measurements of the force exerted on the atom, one could in principle discern the existence of short-range modifications to Newtonian gravity. This avoids the need for a patterned surface where calculations are much harder and for which the probe must be moved laterally at a fixed distance. We also briefly discuss the case where an atomic cloud undergoes Bloch oscillations within an optical lattice created by reflecting a laser off the shield. We find that our scheme has the potential to improve current constraints if relatively modest improvements in atom localisation in optical lattices are made. 
\end{abstract}

%
%
%
%
%

Modifications to the Newtonian gravitational interaction in the submillimeter regime are predicted by a wide range of theories. These include the massive `moduli' fields of string theory whose values determine the geometry of possible extra dimensions \cite{Dimopoulos1996,Antoniadis1998}. An experimental verification of such predictions would be of great significance, but at the same time poses severe challenges. Chief among these is the dominance of electromagnetic interactions in any realistic measurement scheme. We must therefore account for electromagnetic interactions to extremely high accuracy if we are to reveal the faint gravitational signal beneath. 

\begin{figure}[t!]
\centering
\includegraphics[width = 0.7\columnwidth]{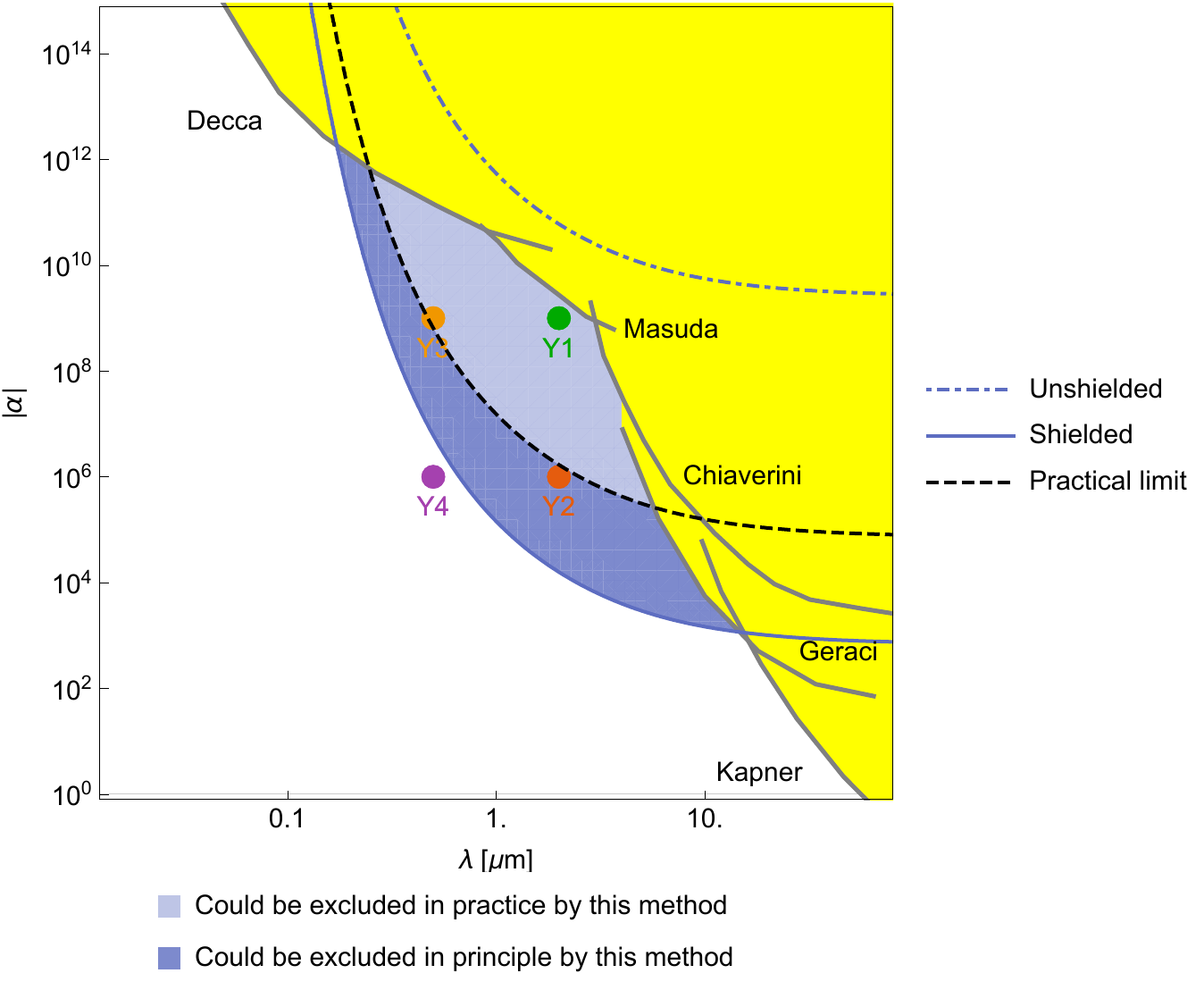}
\caption{Regions of the $\{ \alpha,\lambda \}$ parameter space that are currently excluded by experiments \cite{Decca2005, Masuda2009,Chiaverini2003,Geraci2008,Kapner2007} are shown in yellow. The dark blue region represents the region that could be excluded in principle by the shielding techniques detailed here, i.e. that for which the Yukawa force can be distinguished from all others with no regard as to its absolute value. The light blue region takes into account the experimental setup described at the end of this work, and represents a more practical limit for the power of the method presented here.  The four labelled circles will be used as reference Yukawa parameters throughout the calculation. }\label{ExclusionBackground}
\end{figure}

Despite these challenges, an on-going experimental effort has succeeded in placing constraints on short range deviations to the inverse square law \cite{Adelberger1991,Decca2005, Masuda2009,Chiaverini2003,Geraci2008,Kapner2007,Safronova2018}.  Motivated by theories which predict force mediators with non-zero mass, leading to interaction potentials which decay exponentially with distance, the gravitational interaction potential of two particles of mass $M$ and $m$ separated by a distance $r$ is often parameterised in the form of a Yukawa potential (see, for example, \cite{Floratos1999})
\begin{equation}
U_\text{Y}(r) = \frac{GMm}{r}(1+\alpha e^{-r/\lambda}).
\end{equation}
Here, $G$ is the Newtonian gravitational constant, $\alpha$ is a dimensionless constant and $\lambda$ is a length describing the range of the interaction. The currently allowed ranges of the parameters $\alpha$ and $\lambda$ are summarized in Fig.\ \ref{ExclusionBackground}.

The electromagnetic interactions that will concern us in this paper are Casimir-type forces that arise between bodies due to the surface-induced modification of the zero point electromagnetic field. More precisely, the Casimir force acts between macroscopic bodies whereas the closely related Casimir-Polder (CP) force is the name given to the force between a macroscopic and a microscopic  body (an atom, molecule, nanosphere etc). These forces are usually attractive and vary as the third or fourth power of the inverse distance, depending on the importance of retardation in the specific system involved.   One way to distinguish their effect from gravity is to take advantage of the fact that they depend solely on the electronic properties of the material, while gravitational forces depend only on the mass distribution. This means that objects patterned with regions that have different densities but similar electronic properties can be used to isolate gravitational interactions in short-range force experiments \cite{Sorrentino2009,Klimchitskaya2017}, while other approaches include using a corrugated surface \cite{Bezerra2014}. However, one of the problems with these methods is that a precise calculation of the Casimir and CP forces for such patterned or structured surfaces is extremely difficult due to inherent non-additivity, and the fabrication process can introduce additional complications such as electrostatic surface potentials at the interfaces between materials. The measurements themselves are also a practical challenge as one has to move the force probe laterally across the different regions while maintaining a precise distance from the surface.

\begin{figure}[t]
\centering
\includegraphics[width = 0.6\columnwidth]{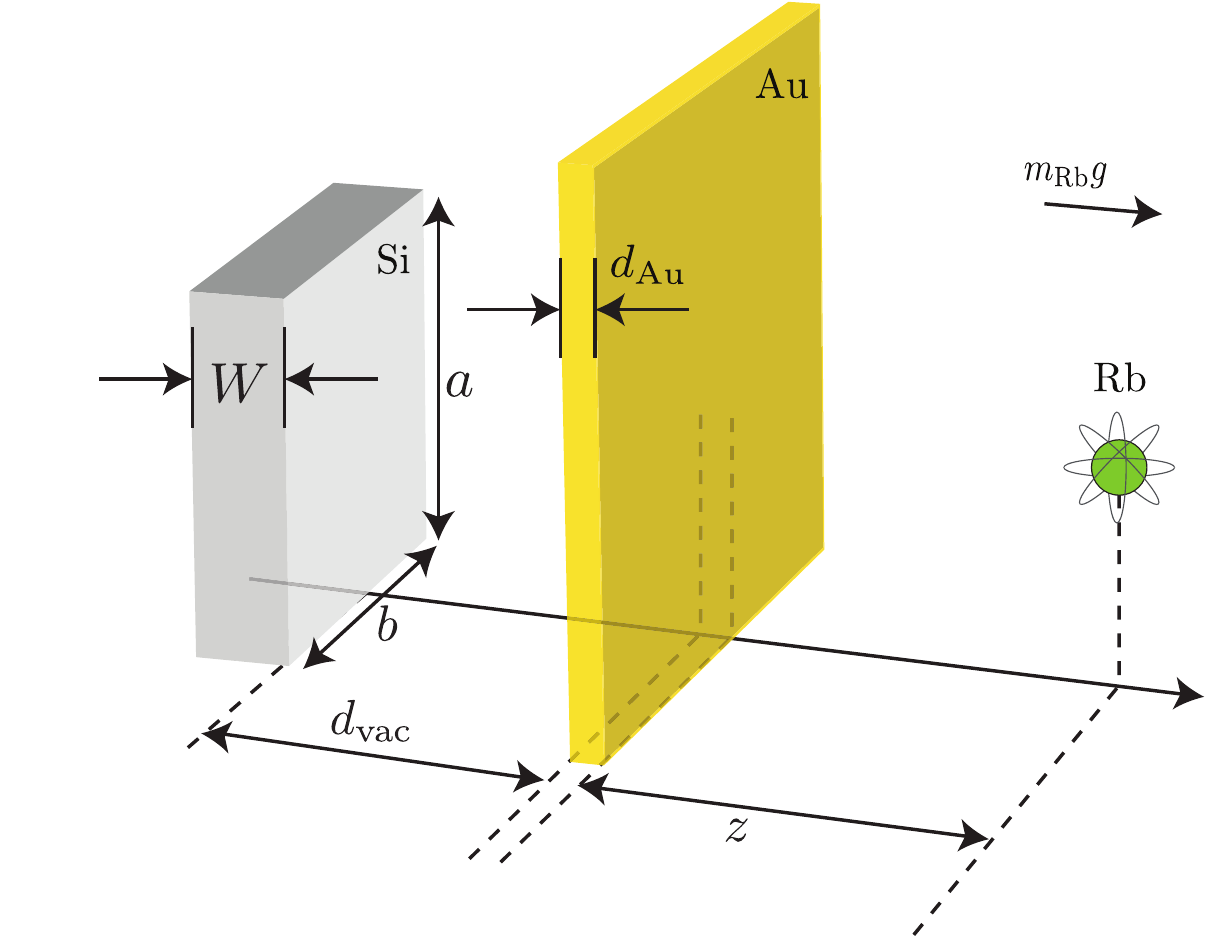}
\caption{Proposed setup for measurement of short-range corrections to Newtonian gravity. The experiment would consist of a silicon slab and a rubidium atom, separated by a gold shield. For reasons that will become clear in Sec.~\ref{MeasurementSchemeSec}, we assume the apparatus is oriented such that the Earth's gravitational force $mg$ is in the direction indicated.}\label{MasterDiagram}
\end{figure} 

Here we take a different approach, avoiding the lateral movement of a probe across a patterned surface and instead rely on the different distance dependences of the hypothetical Yukawa force, Newtonian gravity and the CP force. We base our discussion on the setup  depicted in Fig.\ \ref{MasterDiagram}, where we consider a rubidium atom positioned near a gold sheet, with a medially movable silicon slab behind that. The role of the gold is to shield the atom from the CP force but we specifically avoid making the assumption of perfect reflectivity. Rather,  at the heart of our paper is a precise calculation of the CP force for a shield of finite thickness and conductivity (we are following in the footsteps of a similar calculation for a graphene shield \cite{Ribeiro2013}). We hope this will guide practitioners in evaluating the viability of  ``simple'' medially layered setups like that shown in Fig.\ \ref{MasterDiagram}, as opposed to laterally structured surfaces (where there is less need for a shielded CP calculation). 

We begin in Section \ref{CPForce} by calculating the CP force in such a system, followed in Section \ref{YukawaSec} by an account of the Yukawa force for a slab geometry.  In Section \ref{CompOfForcesSec} we compare the calculated forces to determine which of them dominates. Finally in Section \ref{MeasurementSchemeSec} we briefly explore the feasibility of an experiment measuring the shift in frequency of a Bloch oscillating atom due to a shielded slab.


\begin{figure}[t]
\centering
\includegraphics[width = 0.7\columnwidth]{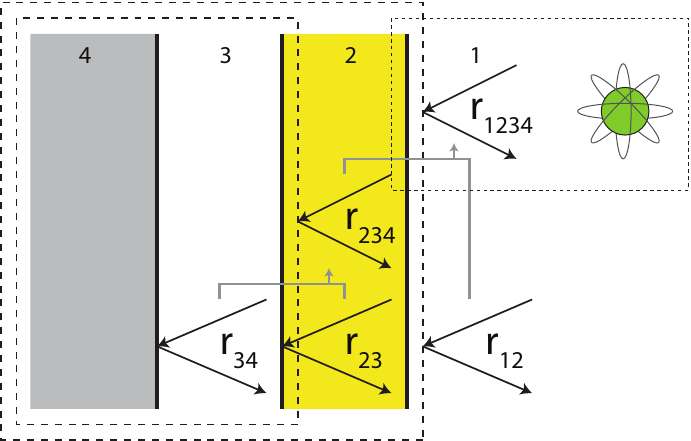}
\caption{Illustration of reflection coefficient nesting process used in our calculations of the CP force.}\label{RCoeffDiagram}
\end{figure}

\section{Casimir-Polder force}\label{CPForce}

Our envisaged experiment features an atom interacting with a multi-layered medium consisting of  vacuum-gold-vacuum-silicon, as shown in Fig.~\ref{RCoeffDiagram}.   In order to have a tractable calculation of the CP force for this situation we will make two simplifying approximations;
\textbf{1)} The silicon slab will be modelled as having infinite depth, and 
\textbf{2)} we assume that all layers are of infinite extent in the lateral directions. 

Let us comment on these simplifications. The first approximation will over-estimate the CP force because as long as the depth of the silicon slab is significantly larger than the atomic transition wavelength (we will choose parameter values satisfying this condition), the CP force from any finite-depth slab will be smaller than that for an infinitely deep one. Note that at the atomic transition wavelength considered here, silicon has quite a large skin depth of $\delta \sim 17\mu$m  \footnote{The skin depth $\delta$ for a non-magnetic material [conductivity $\sigma$ and relative permittivity $\varepsilon(\omega)$] at frequencies much larger than $\varepsilon(\omega) \varepsilon_0/\sigma$ is given by $\delta \sim 2\sqrt{\varepsilon_0 \varepsilon(\omega)/\mu_0}/\sigma$.} so the over-estimation can be quite large for thin slabs. 
The second approximation, meanwhile, is accurate if the lateral dimensions of the apparatus are much larger than all other length scales. Like the first approximation, assuming infinite lateral size will over-estimate the magnitude of the CP force.  Therefore, our approximation scheme will give a reliable \emph{upper bound} for the CP force. Since the CP force represents an unwanted effect when measuring gravity, an upper bound is still of value. By contrast, we shall \emph{not} make these approximations in our calculation of the Yukawa force to be described in Section \ref{YukawaSec}.

The CP potential arises from the interaction energy between a fluctuating dipole and a nearby macroscopic body. The coupling strengths involved are the polarisability of the atom $\alpha(\omega)$ and the reflectivity of the macroscopic body, which for a planar surface is encoded in the TE (transverse electric) and TM (transverse magnetic) reflection coefficients $r_\text{TE}$ and $r_\text{TM}$. The interaction is mediated by photons of wave vector $\B{k}$, which in a planar system is decomposed into parallel ($k_\parallel$) and perpendicular components
\begin{equation}\label{kPerpDef}
k_\perp=\sqrt{(\omega/c)^2-k_\parallel^2} .
\end{equation}
Summing over all such photon contributions entails integrating over any two of the three variables $\omega,k_\parallel,k_\perp$, since the third is fixed by Eq.~(\ref{kPerpDef}). Here we choose to eliminate $k_\perp$. The final step is to rotate the $\omega$-integration to imaginary frequencies $\omega \to i\xi$ ($\xi>0$) in order to avoid a rapidly oscillating integrand. Putting this all together one finds the Casimir-Polder potential $U_\text{CP}$ a distance $z$ from a planar surface \cite{Wylie1984,Buhmann2004,Buhmann2012a}
\begin{eqnarray} \label{MainCP}
U_\text{CP}(z) = \frac{\hbar \mu_0}{8 \pi^2} \int_0^\infty \!\!\!d\xi \, \xi^2 \alpha(i\xi) &\int_0^\infty \!\!\!dk_\parallel  \frac{k_\parallel}{\sqrt{k_\parallel^2+\xi^2/c^2}}  \nonumber\\  &\times\left[  r^\text{TE} - \left(1+2 \frac{k_\parallel^2 c^2}{\xi^2}\right)r^\text{TM} \right] e^{-2z \sqrt{k_\parallel^2+\xi^2/c^2}}
\end{eqnarray}
where the polarizability $\alpha(\omega)$ is defined for a transition $i\to j$ of an isotropically polarizable atom as:
\begin{equation}
\alpha(\omega) = \frac{2}{3\hbar } \lim_{\epsilon\to0}\frac{\omega_{ij}|\mu_{ij}|^2}{\omega_{ij}^2-\omega^2-i\omega\epsilon} \ .
\end{equation}
Here, $\mu_{ij}$ is the atomic transition dipole moment and $\omega_{ij}$ is the transition frequency.

We can build the explicit reflection coefficients for either polarisation $\pol=\, $TE, TM by beginning from the well-known expressions for the overall reflection coefficient  $r^\pol_{ijk}$ of a three-layer system whose two interfaces are separated by a distance $d_j$ and have single-interface Fresnel reflection coefficients $r^\pol_{ij}$ and $r^\pol_{jk}$, and single-interface transmission coefficients $t^\pol_{ij}$ and $t^\pol_{jk}$ as listed in \ref{RandTAppendix}. The composite reflection coefficients are given by \cite{Tomas1995}
\begin{equation}
r^\pol_{ijk} = r^\pol_{ij} + \frac{t^\pol_{ij} t^\pol_{ji} r^\pol_{jk} e^{2 i \beta_j d_j}}{1-r^\pol_{ji}r^\pol_{jk} e^{2 i \beta_j d_j}}
\end{equation}
where $\beta_i =\sqrt{\varepsilon_i \omega^2/c^2-k_\parallel^2}$ is the $z$-component of the wave vector in layer $i$. In our particular system, we have;
\begin{eqnarray}
\beta_1 &= \beta_3 = \beta_\text{vac} = \sqrt{\omega^2/c^2-k_\parallel^2} \\
\beta_2 &= \beta_\text{Au} =\sqrt{\varepsilon_\text{Au}\omega^2/c^2-k_\parallel^2}\\
\beta_4 &= \beta_\text{Si} =\sqrt{\varepsilon_\text{Si}\omega^2/c^2-k_\parallel^2}\\
 \beta_\text{Si} &=\sqrt{\varepsilon_\text{Si}\omega^2/c^2-k_\parallel^2}\\
 d_1 &=z, \quad d_2 = d_\text{Au}, \quad d_3 = d_\text{vac}, \quad  d_4 \to \infty
\end{eqnarray}
where $\varepsilon_\text{Au}$ and $\varepsilon_\text{Si}$ are the relative permittivities of gold and silicon, respectively. This means that $r_{234}$ as shown in Fig.~\ref{RCoeffDiagram} is given by
\begin{equation}
r^\pol_{234} = r^\pol_{23} + \frac{t^\pol_{23} t^\pol_{32} r^\pol_{34} e^{2 i \beta_\text{vac} d_\text{vac}}}{1-r^\pol_{32}r^\pol_{34} e^{2 i \beta_\text{vac} d_\text{vac}}}  \ .
\end{equation}

Now we can build the reflection coefficient of the whole system by composing $r_{234}$ with $r_{12}$:
\begin{equation} \label{r1234}
r^\pol_{1234} = r^\pol_{12} + \frac{t^\pol_{12} t^\pol_{21} r^\pol_{234} e^{2 i \beta_\text{Au} d_\text{Au}}}{1-r^\pol_{21}r^\pol_{234} e^{2 i  \beta_\text{Au} d_\text{Au}}} \ .
\end{equation}
The explicit form of these composite reflection coefficients can be found in \ref{RandTAppendix}. The CP potential of the atom in our particular setup is then finally given by
\begin{eqnarray} \label{CPIntegral}
U_\text{CP} = \frac{\hbar \mu_0}{8 \pi^2} \int_0^\infty \!\!\!d\xi \, \xi^2 \alpha(i\xi) &\int_0^\infty \!\!\!dk_\parallel  \frac{k_\parallel}{\sqrt{k_\parallel^2+\xi^2/c^2}}  \nonumber\\  &\times\left[  r_{1234}^\text{TE} - \left(1+2 \frac{k_\parallel^2 c^2}{\xi^2}\right)r_{1234}^\text{TM} \right] e^{-2z \sqrt{k_\parallel^2+\xi^2/c^2}} \ . \
\end{eqnarray}
This expression allows us to include the effects of finite thickness and conductivity of the shield on the CP force. Note that if the gold layer was a perfect reflector we could set $r_{1234}^\text{TE} \to -1$ and $r_{1234}^\text{TM} \to 1$. 

\begin{figure}[t!]
\centering
\includegraphics[width = 0.6\columnwidth]{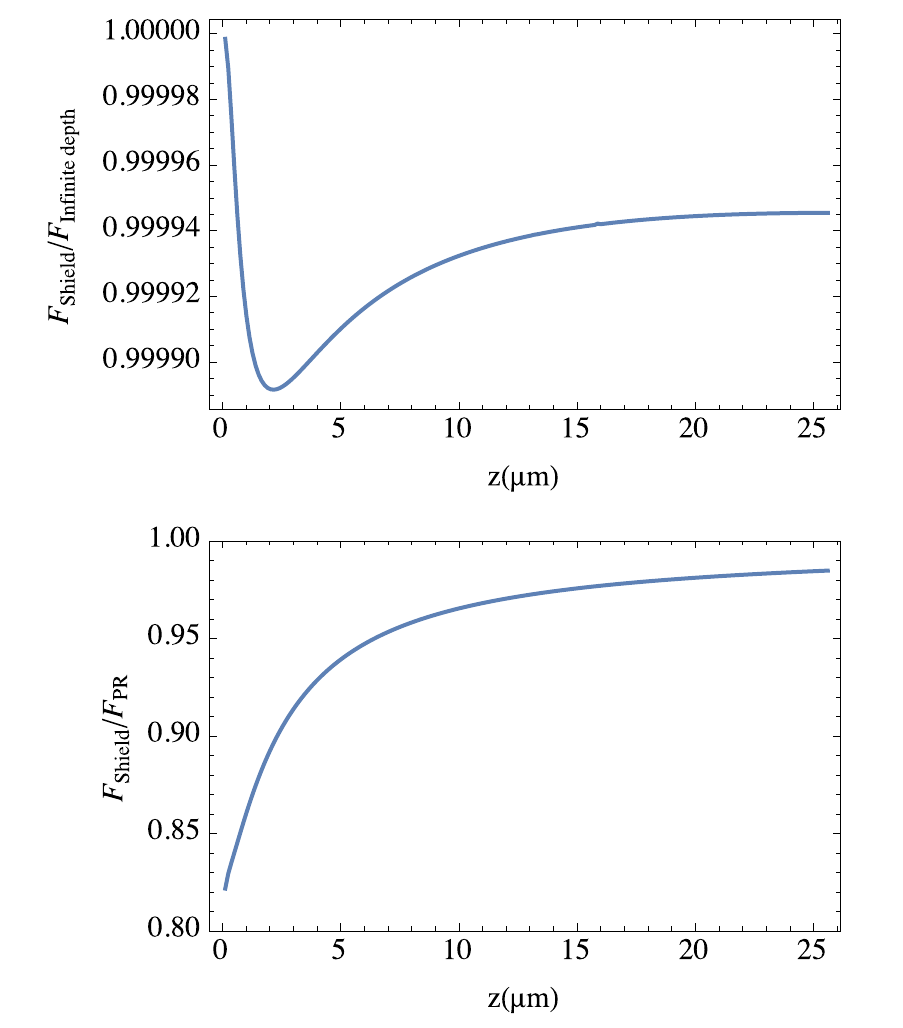}
\caption{\textbf{Upper:} Ratio of CP force near an isolated 50 nm thick sheet of gold to the CP force due to an infinitely thick `slab' of gold. The asymptotic value for large $z$ is not quite equal to 1 because in the far-field the atom can no longer probe the thickness of the shield, resulting in the same distance dependence but simply with a `weakened' infinitely deep dielectric. Nevertheless, it is seen that for thicknesses of 50 nm (or more) the finite thickness of the gold makes almost no difference to the CP force felt near it. \textbf{Lower:} Ratio of CP force near the same $50$ nm thick sheet of gold to that due to a perfectly reflecting surface (whose thickness is of course irrelevant). This, combined with the upper plot, shows that the finite conductivity of the gold has a small but appreciable affect on the CP potential near it.  }\label{CasimirPolderGraphs}
\end{figure} 

The main contribution to the CP potential is at the dominant transition frequency of the atom, which for ground state Rb atoms  is approximately $3.8 \times 10^{14}$Hz ($780$nm). We propose choosing a gold shield of thickness  $d_\text{Au}=50$nm which is considerably greater than the skin depth at this frequency which is of the order of a few nanometers. As shown in the upper plot of Fig.~\ref{CasimirPolderGraphs}, it is indeed the case that  by the time the gold is $50$ nm thick the CP force is rather insensitive to its finite thickness. However, as well as having a finite thickness the gold also has a finite conductivity. This is well-described by the Drude model whose dielectric function is:
\begin{equation}
\varepsilon_\text{Au}(\omega) = 1-\frac{\omega_p^2}{\omega(\omega+i \gamma)}
\end{equation}
where $\omega_p= 1.38\times 10^{16}$rad/s is the plasma frequency and $\gamma \approx 4.08\times 10^{13}$rad/s is the damping parameter \cite{Ordal1985}. This finite conductivity has an appreciable effect on the CP force near the sheet of gold (as compared to a perfect reflector), as shown in the lower panel of Fig.~\ref{CasimirPolderGraphs}. This suggests that in order to get reliable results for the relative values of the various forces involved a realistic CP calculation, as performed here, is necessary.

\section{Yukawa force}\label{YukawaSec}

We now proceed to a calculation of the Yukawa force. This is much simpler than that of the CP force and  we therefore include both the finite depth of the silicon slab and the finite lateral sizes of each layer.  Referring back to Fig.\ \ref{MasterDiagram}, the values we shall use for the various parameters are $a = 100 \mu$m, $b = 100 \mu$m and $W = 10 \mu$m, and as above we assume the thickness of the gold sheet to be $d_\text{Au} = 50$ nm.

\begin{figure}[t]
\centering
\includegraphics[width =0.6 \columnwidth]{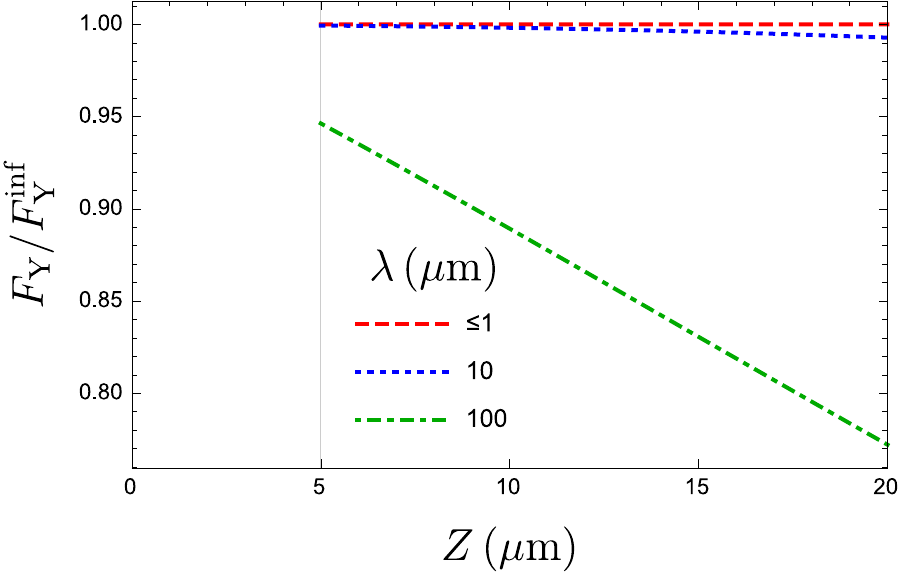}
\caption{The perpendicular component $F_{\text{Y}}$ of the Yukawa force at a distance $Z$ from our finite slab of size $100\mu \mathrm{m} \times 100 \mu \mathrm{m}  \times 10 \mu \mathrm{m}  $ found by numerical integration of Eq.~(\ref{YukawaForceOfArbitraryVolume}) using the density of silicon ($\rho_{\text{Si}}=2330\text{kg/m}^3$). We have expressed the result in units of the Yukawa force $F^\text{inf}_{\text{Y}}$ for a slab of infinite lateral extent, as given by Eq.~(\ref{YukawaForceOfInfiniteSlab}). This ratio is of course independent of the common scaling factor $Gm\rho\alpha$ in eqs~(\ref{YukawaForceOfArbitraryVolume}) and (\ref{YukawaForceOfInfiniteSlab}), but depends on the particular value of $\lambda$ chosen. For $\lambda \leq 1 \mu$m the deviation from unity is not visible at the scale of this graph, so the upper line can be taken to represent all such values of $\lambda$. For larger values of $\lambda$ the force has a longer range, so it makes physical sense that finite size effects should be more visible as $\lambda$ is made larger, as reflected in the figure.}\label{CuboidCorrection}
\end{figure}

In order to calculate the Yukawa force in the system we require an expression for it at a distance $Z = d_\text{vac}+d_\text{Au}+z+W/2$ from the centre of a rectangular slab, as shown in Fig.~\ref{MasterDiagram}. The force is evaluated on the $xy$ symmetry axis of the slab so symmetry dictates that the force $\B{F}_\text{Y}$ is in the $\B{z}$ direction; $\B{F}_\text{Y} = \left[\frac{d}{dZ} U_Y(r)\right] \hat{\B{Z}}=\left[\frac{d}{dz} U_Y(r)\right] \hat{\B{z}}$. We can then integrate this over the slab volume to obtain the Yukawa force from the whole slab. We have
\begin{equation}\label{YukawaForceOfArbitraryVolume}
\B{F}_\text{Y} =(Z\hat{\B{z}}-\B{r}) \frac{G m \rho \alpha}{\lambda} \int_V d^3\B{r} \frac{|Z\hat{\B{z}}-\B{r}|+\lambda}{ |Z\hat{\B{z}}-\B{r}|^3} e^{-|\B{r}-Z\hat{\B{z}}|/\lambda}
\end{equation}
where $V$ is the volume of the slab and $\rho$ is its density. Taking initially the case of a slab of infinite extent in the $xy$ plane and of thickness $W$, we find, in agreement with \cite{Decca2009}, an exact result for the Yukawa force in such a situation
\begin{eqnarray}\label{YukawaForceOfInfiniteSlab}
\B{F}^\text{inf}_\text{Y} &= 4 \pi  \alpha  G \lambda  m \rho e^{-Z/\lambda} \sinh\left(\frac{W}{2\lambda}\right)\hat{\B{z}}
\end{eqnarray}
For a finite slab the integrals must be carried out numerically, which is complicated by the fact that the parameters $\alpha$ and $\lambda$ are unknown. One can eliminate the overall scale $\alpha$ by expressing the finite slab force in units of the infinite slab force, but the range parameter $\lambda$ remains. The results for various values of $\lambda$ of a finite-slab numerical integration are shown in Fig.~\ref{CuboidCorrection}

\section{Comparison of forces}\label{CompOfForcesSec}

\begin{figure*}[t]
\centering
\includegraphics[width = \columnwidth]{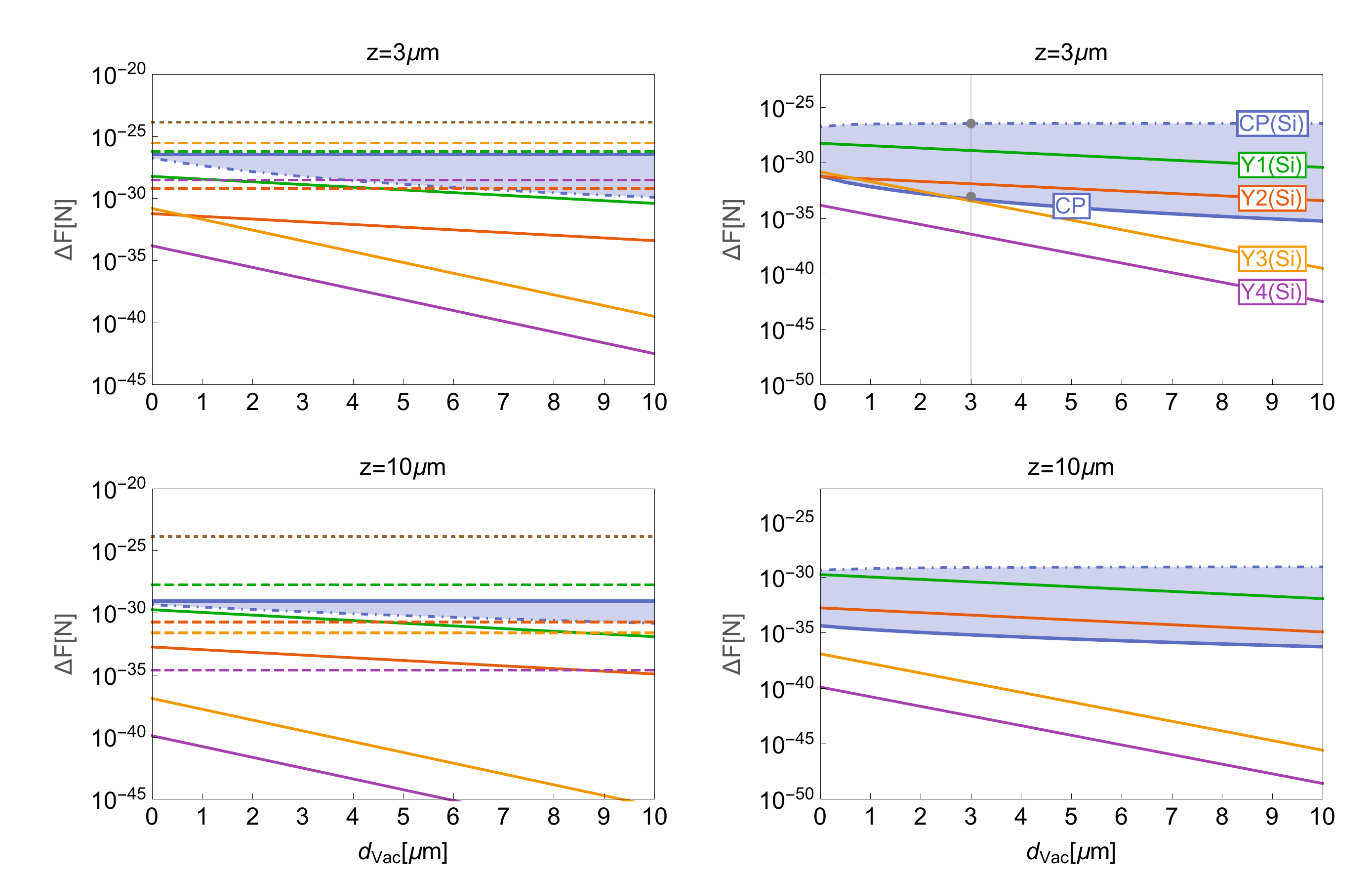}
\includegraphics[width = 0.8\columnwidth]{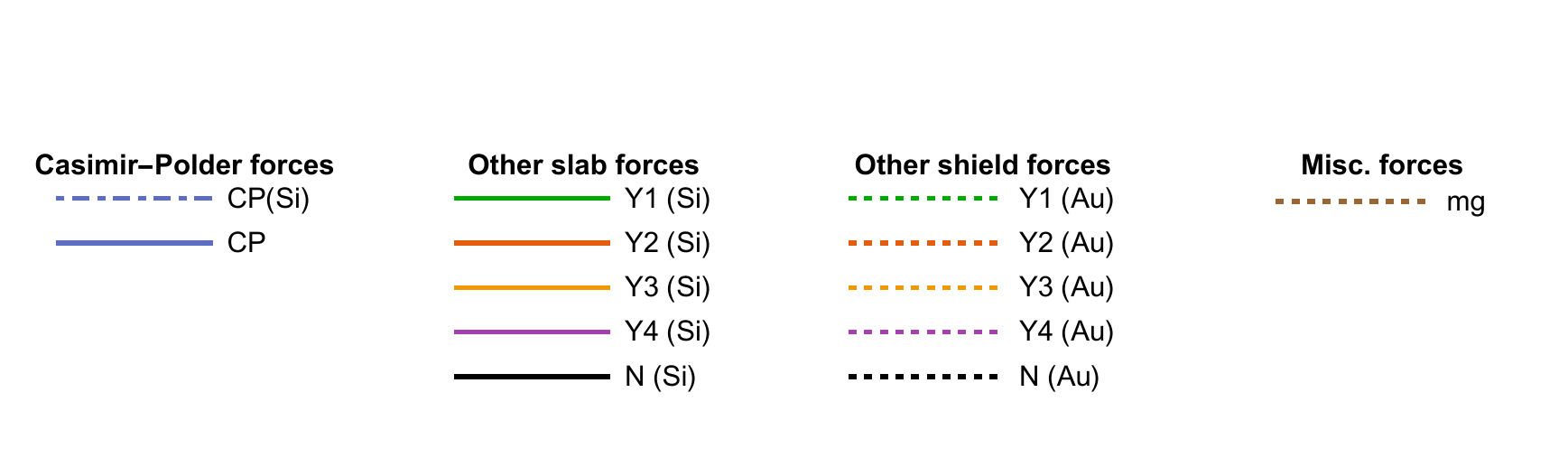}
\caption{Comparison of forces as a function of $d_{\mathrm{vac}}$ at two different fixed distances $3\mu$m and $10\mu$m from the shield, with the parameter choices for the Yukawa force being the ones shown in Fig.~\ref{ExclusionBackground}. The left-hand column of graphs gives the absolute value of the force involved, while the right-hand column gives $\Delta F$ as defined in Eq.~(\ref{DeltaF}). The vertical line and points are for later reference and the shading gives the improvement due to the shield. All the different forces and parameters used are listed in Tables \ref{tab:label} and \ref{ParamsTable}, respectively.}\label{SixGraphsLabelled}
\end{figure*}

Now that we have an account of the Yukawa and CP forces involved in the proposed setup, we can compare them to determine which is dominant and whether any are amenable to measurement. The forces due to each component of the apparatus (and also the Earth) are listed in Table \ref{tab:label} and are plotted in Fig.~\ref{SixGraphsLabelled}  as a function of $d_{\mathrm{vac}}$ (the two rows in Fig.~\ref{SixGraphsLabelled} are for two different fixed values of $z$). The parameters we use are given in Table \ref{ParamsTable} and correspond to the four points chosen in Fig.~\ref{ExclusionBackground}. 
In the plot we have graphed both the absolute values of the forces and the quantity 
\begin{equation}\label{DeltaF}
\Delta F= F -F(d_\text{Vac} \to \infty) 
\end{equation}
so that, for example, the $\Delta F$ for the Yukawa force of the silicon slab is
\begin{equation}\label{FYSi}
\Delta F_\text{Y(Si)}= F_\text{Y(Si)} -F_\text{Y(Si)}(d_\text{Vac} \to \infty) =  F_\text{Y(Si)} 
\end{equation}
and for the gold shield
\begin{equation}
\Delta F_\text{Y(Au)}= F_\text{Y(Au)} -F_\text{Y(Au)}(d_\text{Vac} \to \infty) = 0
\end{equation}
and so on. By subtracting out the value at $d_{\mathrm{vac}}=\infty$, $\Delta F$ gives us the difference between the cases where the silicon slab is present and when it is removed.
\begin{table}[t]
  \centering
  \begin{tabular}{@{} l|p{8cm}|p{5cm}@{}}
    
    Symbol & Description & Force found from \\ 
    \hline 
    Y$_i$(Si) & Yukawa force of the slab with parameters $i=1,2,3,4$ defined in Fig.~\ref{ExclusionBackground}. &$ 4 \pi  \alpha_i  G \lambda  m_\text{Rb} \rho_{\text{Si}} e^{-Z/\lambda_i} \sinh\left(\frac{W}{2\lambda_i}\right)$\\  \hline
    Y$_i$(Au) & As above, but for the shield. &$ 4 \pi  \alpha_i  G \lambda_i  m_\text{Rb} \rho_{\text{Au}} e^{-z/\lambda_i} \sinh\left(\frac{d_\text{Au}}{2\lambda_i}\right)$  \\ \hline
    CP & Casimir-Polder force in the presence of the gold shield, including non-additive contributions from both the slab and the shield. & Numerical evaluation of Eq.~(\ref{MainCP})\\ \hline
    CP(Si) & Casimir-Polder force if the shield is removed.&Numerical evaluation of Eq.~(\ref{MainCP}) with $\varepsilon_\text{Au}(\omega)=1$ \\ \hline
    N(Si) & Newtonian gravitational force of the slab.& $GabW\rho_\text{Si}m_\text{Rb}/Z^2$ \\ \hline
        N(Au) & Newtonian gravitational force of the shield.& $2\pi G \rho_\text{Au} m_\text{Rb}$ \\\hline
    E & Gravitational force of the Earth.& $m_\text{Rb} g$ \\ 
    
  \end{tabular}
  \caption{Summary of forces involved. Here $Z = d_\text{vac}+d_\text{Au}+z+W/2$.}
  \label{tab:label}
\end{table}

\begin{table}
  \centering
  \begin{tabular}{@{} l l | l l |  l l | l l @{}}
    \hline
 \multicolumn{2}{l}{Yukawa } &  \multicolumn{2}{l}{Slab } &  \multicolumn{2}{l}{Shield } &  \multicolumn{2}{l}{Atom}  \\ \hline
    \hline
   $ \{\alpha,\lambda \}_1 $ & $\{10^9,2 \,\mu \mathrm{m} \} $& $\rho_\text{Si}$ & 2 330 kg/m$^3$ & $\rho_\text{Au}$ & 19 300 kg/m$^3$& $\mu_{ij}$ & $5.05 \times 10^{-29}$Cm \\ 
$ \{\alpha,\lambda \}_2 $  & $\{10^6,2 \,\mu \mathrm{m} \} $ & $\varepsilon_\text{Si}$ & 5 &  $\omega_p$ & $1.38\times 10^{16}$rad/s& $\omega_{ij}$  & $2.4\times 10^{15}$rad/s \\ 
$ \{\alpha,\lambda \}_3 $   & $\{10^9,0.5 \, \mu \mathrm{m} \} $ & W & 10 $\mu$m& $\gamma$ & $ 4\times 10^{13}$rad/s& $m_\text{Rb}$&  $1.4 \times 10^{-25}$kg\\ 
$ \{\alpha,\lambda \}_4 $  & $\{10^6,0.5 \,\mu \mathrm{m} \} $ & a,b & 100 $\mu$m &$d_\text{Au}$  & 50 {nm} & &  \\ 
    \hline
  \end{tabular}
  \caption{Summary of parameters used in this work}
  \label{ParamsTable}
\end{table}
 From the left-hand column of Fig.~\ref{SixGraphsLabelled} we see that the Newtonian gravitational force of the Earth ($m_\text{Rb} \, g$) dominates all other forces in all cases, as one would expect. The next most dominant force is the shielded CP force, followed by the Yukawa force corresponding to point $1$ on Fig.\ \ref{ExclusionBackground} which is, however, orders of magnitude smaller. By contrast, we see from the right column of Fig.~\ref{SixGraphsLabelled}, where $\Delta F$ is plotted, that the change in the CP contribution with and without the silicon slab is orders of magnitude less than the change in the Yukawa force corresponding to points $1$ and $2$ on Fig.~\ref{ExclusionBackground}. The Earth's gravity and the various shield forces obviously do not change before and after the removal of the slab, and so do not appear on the right-hand graphs. The figure also emphasizes the change in the CP force due to the presence of the shield, with the improvement to the shielded case highlighted as the shaded area. It is clear from the graphs in the right-hand column that the consideration of realistic shielding of Casimir-Polder forces as given in this paper is necessary to properly evaluate the forces in the proposed system. 

 The key message conveyed by Fig.\ \ref{SixGraphsLabelled} is that if the Yukawa modification to Newtonian gravity exists with any of the parameters chosen in Fig.~\ref{ExclusionBackground}, then the detected change in the total force $\Delta F$ with and without the silicon slab would be much larger than would be expected based on the CP force and Newtonian gravity alone, indicating the presence of a new force. This does not so far say anything about whether the size of that discrepancy is actually detectable; this aspect will be considered in Section \ref{MeasurementSchemeSec}. 

We also briefly consider thermal effects. The regime in which thermal effects can become important to the CP forces is when the black-body peak is at a wavelength comparable to the atom-surface distance (see, for example, \cite{Obrecht2007}). At 300K, this peak is at around $17\mu$m, so for parameters used in the uppermost graphs in Fig.~\ref{SixGraphsLabelled} (which turn out to be the ones we are mainly interested in) we can safely ignore thermal effects even at room temperature. In order to go to longer distances and still be able to ignore thermal effects one would have to cool the apparatus, but cryogenic temperatures are not necessary. For example, at 150K the black-body peak is at approximately $34\mu$m, which easily encompasses the distances we are interested in.

\section{Measurement scheme} \label{MeasurementSchemeSec}

Although our calculations of the forces on an atom detailed above are independent of the measurement method employed, in order to evaluate the viability of the entire scheme presented here let us now examine one particular  method, namely, gravity-induced Bloch oscillations \cite{Sorrentino2009, Poli2011}. This is a type of atom interferometry and takes advantage of the quantum wave-like properties of atoms. However, unlike standard atom interferometers where interference takes place between different paths in coordinate space, and which have, for example, been used to make high precision measurements of the gravitational constant $G$ \cite{Fixler2007,Rosi2014}, Bloch oscillations are the result of interference in momentum space  and the motion of the atoms in coordinate space can be made very small. This allows the atoms to be localised at an almost fixed distance from a surface and is well suited to measuring short-range forces \cite{Carusotto2005}. Indeed, referring to the right hand column of Fig.~\ref{SixGraphsLabelled}, we see that the Yukawa force diminishes significantly with distance (although it should be noted that $\Delta F_{Y}$ remains dominant all the way out to $30\mu$m). Thus, we want the atoms to be located as close as possible to the shield.

Bloch oscillations occur when an external force is applied to a quantum particle that also experiences a periodic potential, which in the present case would be provided by an optical lattice formed by retro-reflection of a laser beam from the gold shield. Bloch oscillations are sensitive directly to the force, as opposed to collective dipole oscillations in a harmonic trap \cite{Harber2005, Antezza2004} (sensitive to force gradients), and so-called super-Bloch oscillations in driven optical lattices (sensitive directly to potential) \cite{Alberti2009}, and can even be measured non-destructively \cite{Peden2009,Venkatesh2009,klesser2016}.  The Bloch oscillation frequency $\nu_B$ is directly proportional to the external force $F$
\begin{equation}\label{BlochEq}
\nu_B = \frac{Fa}{2\pi\hbar}
\end{equation}
 where $a$ is the lattice spacing, which we will take to be $500$nm, but is independent of the depth of the lattice potential. In our proposed experiment the strongest force by some orders of magnitude is $m_\text{Rb}g\approx 1.4 \times 10^{-24}$N, which corresponds to a Bloch oscillation frequency of around 1 kHz. This fast oscillation can be used to our advantage \cite{Carusotto2005}: orienting the apparatus vertically the ``little $g$'' driven oscillations provide a reference oscillator whose frequency can be measured  to an accuracy of one part in $10^7$ \cite{Sorrentino2009, Poli2011}. Moving the silicon slab to different positions, $Z_{i}$ and $Z_{f}$ say, one can attempt to measure the shift $ \Delta \nu_B$ in the Bloch oscillation frequency
\begin{equation}\label{BlochDifferenceEq}
\Delta \nu_B = \frac{[F(Z_f) - F(Z_i)]}{2\pi\hbar} a.
\end{equation}
The difference in the force for the  two positions of the silicon slab is therefore measurable providing $\Delta \nu_B $  is larger than around $10^{-7}\text{kHz}=0.1\text{mHz}$. 

\begin{figure}[t]
\centering
\includegraphics[width = 0.7\columnwidth]{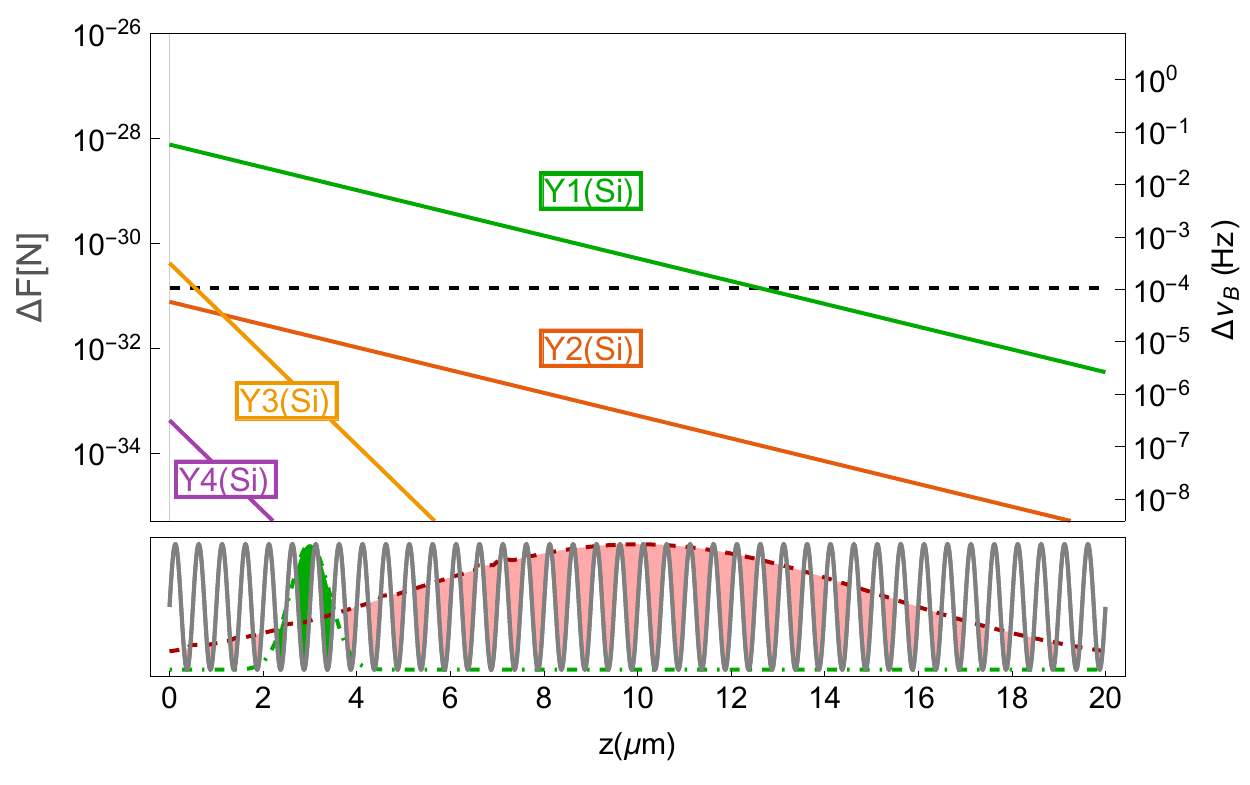}
\caption{Change in Yukawa force (with parameters given by the four points Y1-Y4 in Fig.~\ref{ExclusionBackground}) as $d_\text{Vac}$ is varied between 2.5$\mu$m and $20\mu$m as a function of distance from the shield. The change in Bloch oscillation frequency corresponding to each change in force [calculated from Eq.~\ref{BlochDifferenceEq}] is indicated on the right-hand axis of the plot, alongside the experimental sensitivity of 0.1mHz calculated from \cite{Sorrentino2009, Poli2011}. In the lower panel we show a schematic representation of a 500nm optical lattice containing a sample centred at $10\mu$m from the shield (red), with an r.m.s.\ width 12$\mu$m. As discussed in the text, an r.m.s.\ width appropriate for this experiment would be $2\mu$m (green).}\label{ForceVsZ}
\end{figure}

As mentioned above, we would like to trap the atoms as close to the shield as possible, but in practical terms there is of course a limit to how close to the surface one can go --- this is set by how tightly the atoms can be trapped and how far they move during one Bloch oscillation. For discussion of the former we note from \cite{Ferrari2006, Sorrentino2009} that a typical atomic cloud used in Bloch oscillation experiments has an r.m.s.\ width of approximately $12\mu$m, meaning that trapping distances much closer than this are unrealistic. For example, as shown in the lower part of Fig.~\ref{ForceVsZ}, 
an atomic cloud trapped at $10\mu$m with an r.m.s.\ width of $12\mu$m has appreciable overlap with the surface (approximately 2.5\% of the atoms would be in the region $z<0$). At such a large distance the techniques proposed here would only result in a modest improvement in the region of Yukawa parameter space that can be excluded, as can be seen in Fig.~\ref{ForceVsZ}
 
 We therefore consider what would happen if, in a future experiment, an atomic cloud could be centred at a distance of $3\mu$m from the surface with an r.m.s.\ width of $2\mu$m. As shown in the lower part of Fig.~\ref{ForceVsZ}, the overlap referred to above would be much smaller (approximately $0.02\%$). To this end we note that in deep lattices, or close to the Mott insulator phase \cite{Greiner2002}, atoms occupy close to a single lattice site. In fact, operating in such regimes may not be necessary because the effect of adding an external force on top of a periodic potential is exactly to localise the atoms in space such that the (localised) Wannier-Stark states become the natural eigenfunctions rather than the (delocalised) Bloch functions \cite{Gluck2002}.  During a Bloch oscillation wave packets explore a spatial region of size $w$, given by \cite{Holthaus2000}
\begin{equation} \label{DistanceTravelled}
w = \frac{W}{2F}
\end{equation}
where $W$ is the width of the first energy band. In the tight-binding approximation and with the depth of the lattice being five times the photon recoil energy $E_\text{R}=\hbar^2 k_\text{L}^2/(2m)$, where $k_{L}$ is the laser wavenumber, one finds that $W\approx 0.26 E_\text{R}$  \cite{Holthaus2000}. For our parameters we find $E_R = 6.13 \times 10^{-30}$J, using this in Eq.~(\ref{DistanceTravelled}), one finds $w\approx 0.6 \mu$m, which is a good deal smaller than the proposed trapping distance of $3\mu$m. This means that the oscillations remain localised around $3\mu$m, which is the region in which we have shown the Yukawa force change can dominate. Furthermore, one need not rely solely on little $g$ to provide the fast Bloch oscillation frequency and localisation: an external magnetic field interacting with the atoms' magnetic moment (the basis of magnetic traps) can be used to boost the external force which would also improve the accuracy of the frequency measurement by increasing the number of oscillations during the measurement time.

Finally, we consider which parts of the parameter space of the Yukawa interaction could be excluded in an experiment such as the one we propose here. Firstly we fix $z$ to $3\mu$m based on the discussion above, giving us a value for $\Delta F_\text{CP}$. Equating this to $\Delta F_\text{Y(Si)}$ as given by  Eq.~(\ref{FYSi}) with unspecified $\alpha$ and $\lambda$, one then has an equation that constrains $\alpha$ and $\lambda$ to a particular set of combinations represented by a curve in the $\alpha$-$\lambda$ plane. The region bounded by this curve contains all the values of $\alpha$ and $\lambda$ for which the Yukawa force change is larger than the CP force change in this experiment. This region is shown in Fig.~\ref{ExclusionBackground}, alongside that for the unshielded case (found by solution of $\Delta F_\text{Y(Si)}=\Delta F_\text{CP(S)}$) and the real experimental sensitivity (found by solution of $\Delta F_\text{Y(Si)}=10^{-31}$N, which is the force corresponding to a frequency sensitivity of 0.1mHz as discussed above).  It is seen that the CP shield is vital if any part of the parameter space is to be excluded by this type of experiment, and that there is a considerably sized new region in which the Yukawa force change  overwhelms all others, without consideration of the absolute magnitude (i.e. detectability) of such a force. Taking into account the real experimental sensitivity of a Bloch oscillation experiment, one still finds that a significant new region of parameter space could be excluded. 

\section{Conclusions}

Here we have demonstrated that  a realistic account of the Casimir-Polder force is an important ingredient in the design of experiments using electromagnetic shielding to measure short-range corrections to Newtonian gravity, and that these considerations are in fact the deciding factor in whether the non-Newtonian force can be dominant over all others in a particular experimental setup. Parameterising this force by a Yukawa potential, we have made an initial investigation into whether this force is large enough to be measurable, finding that, given modest improvements in localisation of atoms in an optical lattice, a new region of the Yukawa parameter space can be excluded. 

\section{Acknowledgements}
We gratefully acknowledge E. A. Hinds, J. D. D. Martin and I. Yavin for advice and suggestions. D.O. thanks the Natural Sciences and Engineering Research Council of Canada (NSERC) for funding, and R.B. thanks the UK Engineering and Physical Sciences Research Council (EPSRC), the Alexander von Humboldt Foundation and Freiburg Institute for Advanced Studies (FRIAS).

\appendix

\section{Reflection and transmission coefficients}\label{RandTAppendix}

The single-interface reflection and transmission coefficients are;
\begin{eqnarray}
r^\text{TE}_{ij} &= \frac{\beta_i - \beta_j}{\beta_i + \beta_j} \qquad   r^\text{TM}_{ij} &= \frac{\varepsilon_j \beta_i -\varepsilon_i \beta_j}{\varepsilon_j \beta_i + \varepsilon_i \beta_j}\\
t^\text{TE}_{ij} &=1+r^\text{TE}_{ij}  \qquad   t^\text{TM}_{ij} &=\frac{\varepsilon_i}{\varepsilon_j} \left(1+r^\text{TM}_{ij} \right)
\end{eqnarray}
Using a shorthand $S_i = e^{2 i \beta_i d_i}$, the four layer reflection coefficient used in the text is, for either polarization;
\begin{equation}
r^\sigma_{1234} = r^\sigma_{12}-\frac{S_2(r^\sigma_{12}+1) (r^\sigma_{21}+1) \Big[S_3(r^\sigma_{32}+1) r^\sigma_{34}+r^\sigma_{23} \left(1+S_3r^\sigma_{34}\right)\Big]}{S_3r^\sigma_{32} r^\sigma_{34}+S_2r^\sigma_{21} \Big[S_3(r^\sigma_{32}+1) r^\sigma_{34}+r^\sigma_{23} \left(1+S_3r^\sigma_{34}\right)\Big]-1}
\end{equation}
which is a more explicit version of Eq.~(\ref{r1234})

\newpage
\section*{References}
\bibliographystyle{unsrt}

\end{document}